\documentclass{aa}
\usepackage{graphics}
\usepackage{times}
\newcommand{\aaa}{A \&\,A }

\newcommand{\aas}{A\&\,AS }

\newcommand{\apj}{ApJ }
\newcommand{\apjs}{ApJS }
\newcommand{\apss}{Ap\&\,SS }
\newcommand{\araa}{ARA\&\,A }

\newcommand{\mn}{MNRAS }

\begin{document}

 \thesaurus{09.03.2, 02.13.1, 10.07.1}
 \title{The Galactic magnetic field and propagation of ultra-high energy
 cosmic rays}
  \author{M.~Prouza and R.~\v{S}m\'{\i}da
  }
  \offprints{prouza@fzu.cz}
  \institute{Center for Particle Physics, Institute of Physics, The
  Academy of Sciences of the Czech Republic, Na Slovance 2,\\
  CZ-182~21~Praha 8, Czech Republic}

\authorrunning{M. Prouza \and R. \v{S}m\'\i da}
\titlerunning{GMF and Propagation of UHECR}

\date{Received ??? ??, 2002, accepted ??? ??, 2003}

\abstract{The puzzle of ultra-high energy cosmic rays (UHECRs)
still remains unresolved. With the progress in preparation of next
generation experiments (AUGER, EUSO, OWL) grows also the
importance of directional analysis of existing and future events.
The Galactic magnetic field (GMF) plays the key role in source
identification even in this energy range. We first analyze current
status of our experimental and theoretical knowledge about GMF and
introduce complex up-to-date model of GMF. Then we present two
examples of simple applications of influence of GMF on UHECR
propagation. Both examples are based on Lorentz equation solution.
The first one is basic directional analysis of the incident
directions of UHECRs and the second one is a simulation of a
change of chemical composition of CRs in the energy range $10^{13}
\div 10^{19}$ eV. The results of these simple analyses are
surprisingly rich --- e.g. the rates of particle escape from the
Galaxy or the amplifications of particle flux in specific
directions. }

\maketitle

\newpage

\section{Introduction}

The origin of the high-energy cosmic rays and the ultra-high
energy cosmic rays (UHECRs\footnote{For the purposes of this
article we define ultra-high energy cosmic rays (UHECRs) as cosmic
rays with energy above $10^{19}$ eV and extremely high energetic
cosmic rays (EHECRs) as cosmic rays with energy above $10^{20}$
eV.}) is one of the major unresolved questions in astrophysics,
with a degree of uncertainty increasing with energy of the
particles. The situation is more complicated than in radio,
optical or TeV gamma-ray astronomy, where we observe arrival
directions of non-charged photons and we can easily locate the
positions of their sources from these observations. However,
because it is generally accepted that the primary particles with
energies above $10^{12}$ eV or significant part of them are fully
ionized and therefore charged atomic nuclei, we must consider the
influence of magnetic fields on their propagation from the source
to the Earth. This deflection prevents unambiguous identification
of possible sources.

It is generally believed that the bulk of CRs with the energy
below the knee (around $3\times10^{15}$ eV) has Galactic origin
and its main production mechanism is an acceleration by supernovae
shocks (\cite{axford}). But the origin of the knee remains a
mystery. CRs with energies above the knee may be explained either
as of extragalactic and or Galactic origin. Since the Larmor radii
of the particles with the energy in EeV region become larger than
the thickness of the Galactic disk, it is likely that their
sources are extragalactic. The interesting aspect of the
extragalactic CRs with energies exceeding 50 EeV are the energy
losses due to the interactions with cosmic microwave background.
These energy losses\footnote{Mean interaction length is about 6
Mpc, energy loss is about 20 \% of actual particle energy per
collision.} constrain detected UHECRs to have been produced in the
sources within 100 Mpc. This distance restriction is known as
Greisen-Zatsepin-Kuzmin (GZK) cutoff (\cite{greisen},
\cite{zacepin}).

Earth's atmosphere absorbs high energy cosmic rays and so they reveal
their existence on the ground only by indirect effects such as
ionization and showers of secondary charged particles covering areas up
to many km$^2$. The energy flux of CRs is rapidly decreasing with their
increasing energy. We observe one particle per m$^2$ per year at
energies of $10^{15}$ eV but only one particle per km$^2$ per year at
energies of $10^{18}$ eV. Thus, we need a large detector to find and measure
these rare events. In the next decade the Pierre Auger Observatory
should be able to collect several hundreds of events above the GZK
cutoff, at least ten times more than all events detected up to now.

We use simple method to model the propagation of cosmic rays in a
wide range of energy (from $10^{13}$ eV to the highest value ever
detected $3.2\times10^{20}$ eV). Although our method --- solution
of Lorentz equation --- is the simplest method of modelling of
propagation of CRs, we show that it can be successfully used for
wide range of different applications. The results of modelling the
directional analysis of UHECR and the chemical composition of
Galactic CRs are presented in this work for one complex model of
GMF. In addition, we discuss experimental evidence about GMF and
other GMF models and we also investigate the influence of
turbulent magnetic fields.

\section{Galactic magnetic field}
\subsection{Experimental evidence}

The first evidence of the existence of a Galactic magnetic field
was derived from the observation of linear polarization of
starlight by \cite{hiltner}. Many new measurements were done since
then using the Zeeman spectral-line splitting (gaseous clouds,
central region of the Galaxy), the optical polarization data
(large-scale structures of the magnetic field in the local spiral
arm) and the Faraday rotation measurements in the radio continuum
emission of pulsars and of the extragalactic sources. The last
mentioned method is probably also the most reliable for the large
scale structure. This method is also used for the determination of
the global structure of the magnetic fields in the external
galaxies. From these measurements it follows that the Galactic
magnetic field has two components --- regular and turbulent
(\cite{rand}).

Random fields appear to have a length scale $50 \div 150$ pc and
they are about two or three times stronger than the regular field.
These random field cells have such a small scale (in comparison
with kiloparsec scale of Larmor radii of UHECRs) that they are not
modelled within global GMF models.  However, it follows from
recent work of \cite{harari} or \cite{jaime} that turbulent field
really plays key role in the clustering, magnification or
multiplying of the source images. Therefore we introduced random
fields into our simulations, respecting the fact that such fields
are very strong especially in Galactic arms regions.

We are able to summarize our direct experimental knowledge about
the Galactic magnetic field in several statements (according to
\cite{beck}, \cite{widrow} and \cite{han_new}):

\begin{itemize}
\item{The strength of the {\em total} magnetic field in the Galaxy
is $(6 \pm 2)$ $\mu$G in the disk and about $(10 \pm 3)$ $\mu$G
within 3 kpc from the Galactic center.} \item{The strength of the
local regular field is $(4 \pm 1)$ $\mu$G. This value is based on
optical and synchrotron polarization measurements. Pulsar rotation
measures give more conservative and approximately twice lower
value. These rotation measures are probably underestimated due to
anticorrelated fluctuations of regular field strength and of
thermal electron intensity. On the other hand, optical and
synchrotron polarization observations could be overestimated due
to presence of anisotropic fields.} \item{The local regular field
may be a part of a Galactic magnetic spiral arm, which lies
between the optical arms.} \item{The global structure of the
Galactic field remains uncertain. However, an established
conservative model, which prevails in the last years, is the
two-arm logarithmic spiral model (see below).} \item{Existence of
two reversals in the direction towards Galactic center was
confirmed recently. The first reversal is lying between the Local
and Sagittarius arm, at $\sim 0.6$ kpc from the Sun, the second
one is lying at $\sim$ 3 kpc from the Sun. Some of the Galactic
reversals may be due to large-scale anisotropic field loops.}
\item{As expected from the beginning of the 1990s and also
recently confirmed, the Galactic center region contains highly
regular magnetic fields with strengths up to 1 mG. This extremely
intensive field is concentrated in thin filaments oriented
perpendicularly to the Galactic plane. The characteristic length
of these filaments is about 0.5 kpc.} \item{The local Galactic
field is oriented mainly parallel to the plane, with a vertical
component of only $B_z \simeq (0.2 \div 0.3)$ $\mu$G in vicinity
of the Sun. The recent explanation is that this component is
present due to existence of poloidal magnetic field (see
theoretical global field model below) --- poloidal field naturally
originates within dynamo model of GMF generation.} \item{The
Galaxy is surrounded by a thick radio disk (height of about 1.5
kpc above and under Galactic plane, half-width of 300 pc) similar
to that of the edge-on spiral galaxies. The field strength in this
thick disk is estimated to be around 1 $\mu$G. As in the case of
vertical field component discussed above (poloidal field), the
most common explanation of existence of such thick disc is that
this field is toroidal field originating through dynamo effect.}
\item{The local Galactic field in the standard thin disk has an
even symmetry with respect to the plane (it is a quadrupole). This
is in the agreement with the galactic dynamo model, which is
briefly discussed in the next paragraph.}
\end{itemize}

Other facts used in modelling of GMF have indirect character ---
they are usually derived from the observations of the other spiral
galaxies and of the structure of their magnetic fields or from
existing hypothesis of the mechanisms of magnetic field
generation. In general, it is expected, that the Galactic magnetic
field encompasses the entire  Galactic disk and shows some spiral
structure. Further research and measurements in this field have
vital importance not only for the observations of UHECRs, but also
for the whole cosmic-ray physics and for other astronomical
applications, e.g. for Galactic dynamics.

\subsection{Theoretical global models of GMF}\label{gmf}

\begin{figure}
\resizebox{\hsize}{!}{\includegraphics{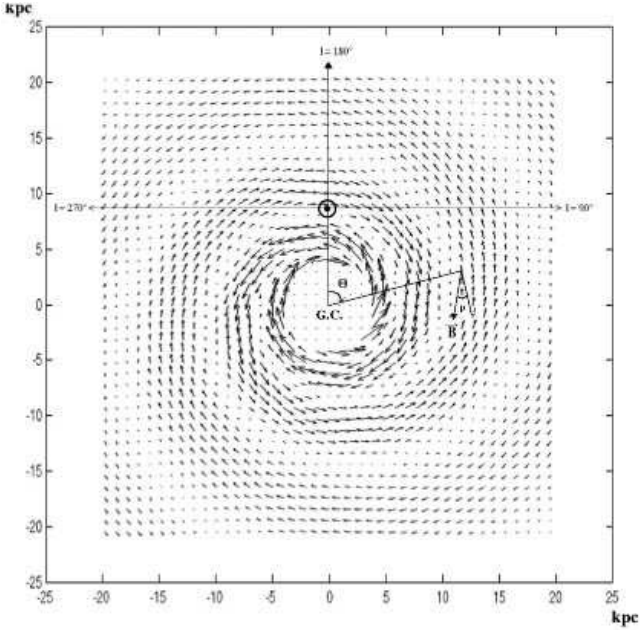}} \caption[
]{\em Direction and strength of the regular magnetic field in the
Galactic plane is represented by the length and direction of the
arrows. The field inside the Galactocentric circle of radius 4 kpc
is taken as constant, 6.4 $\mu$G. The model was constructed using
parameters from \cite{stanev1} and \cite{han}, which are in detail
described in the text. The sense and orientation of the field and
of the angles $\Theta$ and $p$ follows from the figure. G.C.
denotes the Galactic center (at $l = 0^{\circ}$).}

\label{modelgmf}
\end{figure}

The global models omit the presence of turbulent fields and they
are trying to model just the regular component. The basic
conservative model of global Galactic plane was established by
\cite{hanold}, based on the Faraday-rotation measurements of 134
pulsars. The model assumes a two-arm logarithmic spiral with the
constant pitch angle\footnote{The pitch angle determines the
orientation of local regular magnetic field. Its sense is clear
from Fig. \ref{modelgmf}. Precise definition of pitch angle is not
unique, in this work we used the definition proposed by
\cite{han}: The galactic azimuthal angle $\Theta$ is defined to be
increasing in the direction of galactic rotation. Logarithmic
spirals are then defined by:
\begin{equation}
R = R_{0} e^{k \Theta},
\end{equation}
where $R$ is the radial distance and $R_0$ is the scale radius.
The pitch angle is then $p = \mbox{arctan}(k)$. This angle is
negative for trailing spirals such as in our Galaxy, where $R$
increases with decreasing azimuthal angle $\Theta$. For our
Galaxy, the galactic angular momentum vector points toward the
south Galactic pole, and $\Theta$ increases in a clockwise
direction when viewed from the north Galactic pole.} $p$ and it
shows $\pi$-symmetry, so that it is bisymmetric (BSS) magnetic
field model. More exactly, it has also a dipole character (it has
field reversals and odd parity with respect to the Galactic
plane), so it is called BSS-A model.

Discussed model employs cylindrical coordinates --- the radial
distance $r$, the position angle $\Theta$ and the vertical height
$z$. The radial $B_r$ and azimuthal $B_{\Theta}$ components at the
plane position $(r, \Theta)$ can be given by the following
equations:
\begin{equation}
B_{\Theta} = B_0(r) \cos \left( \Theta - \beta \ln \frac{r}{r_0}\right)
\cos p,
\end{equation}
\begin{equation}
B_{r} = B_0(r) \cos \left( \Theta - \beta \ln \frac{r}{r_0}\right) \sin
p.
\end{equation}
where $p$ denotes the pitch angle and according to \cite{stanev1}
and \cite{han_new} it is about $-10^{\circ}$, $\beta = 1/ \tan p
\doteq -5.67$, $r_0$ is the Galactocentric distance of the maximum
field strength at $l = 0^{\circ}$ (in the discussed model it has a
value $r_0 = 9.0$ kpc) and for $B_0(r)$ it holds:
\begin{equation}
B_0(r) = 3 \frac{R}{r},
\end{equation}
where $R$ is the Galactocentric distance of the Sun, taken as 8.5 kpc.

The vertical ($z$) component of the field is taken as zero in approximate
agreement with observations. Results of this model are
depicted on Fig. \ref{modelgmf} and the orientation of the whole system
is also clear from this Figure.

The size and field strength in the Galactic halo is extremely important
for the cosmic-ray trajectories, but it is very poorly known, as we stated
above. Obvious approach to this problem is represented by the work of
\cite{stanev1}, where the field above and under the Galactic
plane is taken as exponentially decreasing:
\begin{equation}
|B(r, \Theta, z)| = |B(r, \Theta)| e^{(-|z|/z_0)},
\end{equation}
where $|B(r, \Theta)|$ is the vector sum of magnitudes of $B_r$
and $B_{\Theta}$ with the $z_0$ = 1 kpc\footnote{There is a slight
difference in comparison with \cite{stanev1}, he used two-scale
model --- with the $z_0$ = 1 kpc for $|z| < 0.5$ kpc and $z_0$ = 4
kpc for $|z| > 0.5$ kpc.}.

We used this described model of GMF as the basis for our
simulations. Strictly speaking, we simply add toroidal and
poloidal field components to this model, as it is detailed in the
next subsection.

Alternative models with another field configurations were also
proposed. The another possible but according to recent
observations a bit less probable configuration is the so-called
ASS-S configuration, axisymmetric configuration without reversals
and with even parity (\cite{stanev1}). However, this configuration
has one advantage. It could be much easier modeled using of the
very popular dynamo model of magnetic field generation
(\cite{elstner}). The bisymmetric mode can also be obtained from
dynamo model, but in such case the use of strong non-axisymmetric
perturbations is necessary. The other two possibilities of
magnetic field configurations --- bisymmetric dipole type (BSS-S)
and axisymmetric quadrupole type (ASS-A) are also not completely
observationally excluded yet (\cite{beckold}).

\subsection{Poloidal and toroidal regular field components}

The dynamo model has one very interesting consequence for the
propagation of CRs --- namely that except of relatively flat field
in the galactic disc it contains also quite strong toroidal fields
above and under the galactic plane. Motions of these fields and
their superpositions generate the net field in the Galaxy. The
existence of such field is indirectly supported by the existence
of radio thick disc mentioned above in the review of observation
results. Such field could change the CR trajectories quite
essentially, furthermore this type of models was not yet used for
UHECR propagation simulation, therefore we decided to add these
components in our simulations. We take advantage from the fact
that only recently some first quantitative estimates of strengths
of such fields were proposed by \cite{han_new}.

For toroidal field we choose the model with simple geometry
(circular discs above and under Galactic plane with Lorentzian
profile in z-axis). For cartesian components of toroidal field it
holds\footnote{The equations above are valid only in the northern
Galactic hemisphere, in the southern hemisphere the field has an
opposite direction, so $B_x$ and $B_y$ components will change
their sign there.}:
\begin{eqnarray}
B_x &=& -B_T \sin(\phi)\\
B_y &=& B_T \cos(\phi)
\end{eqnarray}
For the value of $B_T$ we have:
\begin{eqnarray}
B_T &=& B_{max} \frac{1}{1+(\frac{z-H}{P})^2}\\
 & & \mbox{ for $x^2+y^2<R^2$ and } \nonumber \\
B_T &=& B_{max} \frac{1}{1+(\frac{z-H}{P})^2}\exp
\left(-\frac{(x^2+y^2)^{1/2}}{R}\right)\\ & & \mbox{ for
$x^2+y^2>R^2$} \nonumber
\end{eqnarray}
where $x$ and $y$ are positions in Galactic plane. Meaning and
values of used constants follow: radius of a circle with toroidal
field $R=15$ kpc, height above Galactic plane $H=1.5$ kpc,
half-width of Lorentzian distribution $P=0.3$ kpc, and maximal
value of toroidal magnetic field $B_{max}=1\mu$G.

As consequence of existence of the poloidal field (dipole field)
we probably observe vertical component of $0.2 \mu$G in the Earth
vicinity and intensive filaments near Galactic center. Appropriate
equations, which we used for description of poloidal field, are
the same as the equations for magnetic dipole. The field is
symmetrical around Galactic axis. For the total poloidal field
strength it is then valid (in xz-plane) in polar coordinates
($\theta$ ranges from 0 to $\pi$ and it goes from north to south
pole):
\begin{equation}
B = \frac{K}{R^3} \sqrt{3\cos^2(\theta)+1} \mbox{.}
\end{equation}
From it follows that in spherical coordinates we then have these
cartesian field components:
\begin{eqnarray}
B_x &=& -\frac{3K}{2R^3}\sin 2\theta \cos \phi\\
B_y &=& -\frac{3K}{2R^3}\sin 2\theta \sin \phi\\
B_z &=& -\frac{K}{R^3} (3 \cos^2 \theta - 1)
\end{eqnarray}

A cylinder (height $300$ pc, diameter $100$ pc) with constant
strength of magnetic field equal to $2$ mG was put into Galactic
center instead of field resulting from equations
above\footnote{Orientation of this field is in accordance with
general description, only the strength is constant.}. Main motive
for such arrangement was to avoid a problem with too strong field
near this center ($R\sim0$) and so to keep total field strength in
observed bounds and to describe character of observed filaments.

\begin{figure}[h]
\resizebox{\hsize}{!}{\includegraphics{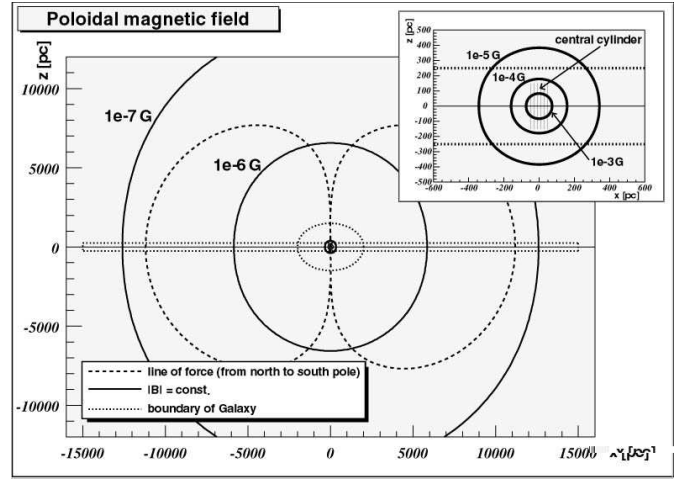}} \caption[ ]
{\em The resulting model of poloidal magnetic field. The central
region is enlarged in the upper right corner.} \label{poloidal}
\end{figure}

The constant $K$ was selected as follows: $K=10^5$ G$.$pc$^3$ for
outer regions ($ R > 5$ kpc) and $K = 200$ G$.$pc$^3$ for central
region ($R < 2$ kpc). For the intermediate region (2 kpc $< R <$ 5
kpc) we used constant absolute field strength $10^{-6}$ G. These
values correspond with observed features of Galactic magnetic
field: milligauss field is restricted only to the central cylinder
and the vertical magnetic field is equal to $0.2$ $\mu$G in the
Sun's distance (see also Fig. \ref{poloidal}).

\subsection{Turbulent fields and Galactic arms}

As we pointed out above, we introduce also the influence of random
fields in our simulations. Cells with characteristic size of 50 pc
with random field orientation and with maximum field strength $|B|
= 10 \mu$G were added to regular, poloidal and toroidal field
components.

Within two following examples we used two different approaches for
introduction of turbulent fields. For study of chemical
composition of CR (Example No. 2) various configurations (cell
frequency, cell size) were used and are in detail described below.
However, for directional analysis (Example No. 1) we respect the
fact, that turbulent fields are common especially within spiral
arms of Galaxy. Therefore we expect that 80\% of volume inside
spiral arm regions contain turbulent field component, while
outside these arm regions only (but within surroundings of
Galactic plane\footnote{More precisely: For the distance $r <$ 20
kpc and $|z| < 1.5$ kpc, where $r$ and $z$ are components of
cylindrical Galactic coordinates.}) 20\% of total simulated volume
have also nonzero turbulent field. Finally, in other outer regions
of the Galaxy we suppose that only 1\% of volume has also nonzero
turbulent component.

\begin{figure}[h]
\resizebox{\hsize}{!}{\includegraphics{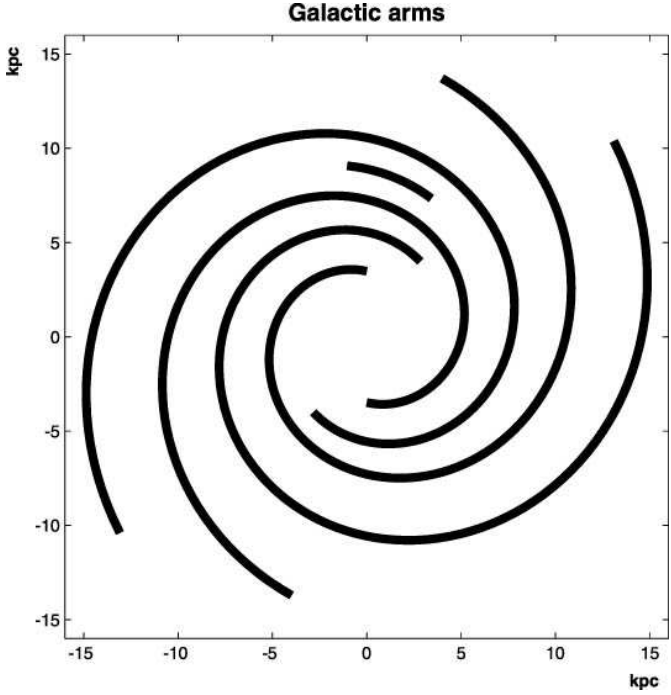}} \caption[
]{\em Model of Galactic arms by \cite{wainscoat} has 4-arm pattern
and it also includes so called Local arm. Equation for individual
arms is: $\Theta(R) = \alpha \log( R/R_{min}) + \Theta_{min}$,
where $\alpha$ is winding constant, $R_{min}$ is inner radius and
$\Theta_{min}$ angle at inner radius (Sun is positioned at
y-axis.). Width of each arm is 750 pc, height 500 pc (centered on
Galactic plane). Arms are truncated at distance 15 kpc from
Galactic center. Parameters for individual arms follow in form
(Arm No., $\alpha$ [rad], $R_{min}$ [kpc], $\Theta_{min}$, angular
extent ($\Theta_{max} - \Theta_{min}$) [rad]): (1, 4.25, 3.48,
1.571, 6.0); (1', 4.25, 3.48, 4.712, 6.0); (2, 4.89, 4.90, 4.096,
6.0); (2', 4.89, 4.90, 0.953, 6.0); (Local, 4.57, 8.10, 1.158,
0.55).} \label{galarms}
\end{figure}

As model of spiral arms we used model by \cite{wainscoat}, which
is simple four-plus-local arm model. Parameters of this particular
model are in detail described in Fig. \ref{galarms}.

\section{Propagation of UHECRs in GMF}

Within the next sections we describe two simple analyses of cosmic ray
propagation in GMF. These analyses are done in different energy ranges
and are serving for derivation of different conclusions, but they are
involving the very same principles of particle motion in magnetic
fields:

The propagation of the main part of UHECR (or more generally of
cosmic rays) candidates (charged particles like protons, nuclei,
electrons, \dots) is of course influenced by the magnetic fields.
This influence is given simply by the well-known Lorentz equation.
The member with electric field in this equation could be
neglected, because there is no evidence for large-scale electric
fields in the Galaxy. For the acceleration $\vec a$ we get then:
\begin{equation}
\vec a = \frac{q}{m} (\vec v \times \vec B),
\end{equation}
where $q$ is the charge of particle, $m$ is its relativistic mass, $v$
its velocity\footnote{Almost equal to velocity of light $c$; UHECRs are
reaching the highest known relativistic $\gamma$-factors, about
$10^{11}$.} and $B$ is the magnetic field strength.

Taking $\vec B$ as constant in suitable small volumes the trajectory
of a particle is followed and the resulting deflection is
examined.

\section{Application No.1: Directional analysis of UHECR}
\subsection{``Antiparticle tracing'' method and recent works}

Some computer simulations in the UHECR range were treated
for this purposes recently and the effects especially on the changes in
spatial distribution were studied.

The method of ``antiparticle tracing'' is used in all these models. The
particle carrying the opposite charge starts its propagation on the
position of the Earth in the Galaxy. Its initial velocity vector has
spherical coordinates $\sim c, b, l$, where $c$ is the velocity of the
light and $b$ and $l$ are the galactic coordinates of the detected
particle arrival. Because of the opposite charge such particle traces
backwards the trajectory of original detected particle. When the particle
leaves from the sphere of influence of the Galactic magnetic field, we
are able to evaluate its new galactic coordinates and thus its initial
direction before the entrance into GMF.

The first work was published by \cite{stanev1}. It analyzes the
motion of UHECRs in conservative models of BSS-A and ASS-S GMF
with similar parameters as were given above. \cite{stanev1}
examines the shifts for protons with energies ranging from $(2
\div 10) \times 10^{19}$ eV. The second article is by
\cite{medina}. The particles with energy equal to $4\times
10^{19}$ eV are analyzed in this paper. The changes in regular
distributions are followed for the ASS-S model of GMF and for the
particles supposed to be either protons or Fe nuclei. The basic
results of both models (magnitudes of deflections) are in good
agreement with our model.

Two other papers appeared recently. In these papers the GMF model of
\cite{stanev1} was employed to support of specific arguments. Firstly,
\cite{olinto} assumed iron nuclei as the only component of UHECRs and
the authors were trying to identify the sources as very young pulsars.
Secondly, \cite{tinyakov} investigated correlation between the positions
of UHECRs propagated outside from Galaxy and of positions of specific
type of blazars. They focused on possible identification of these
blazars as UHECR sources and significant attention was payed also to
analysis of clustered UHECR events.

Two other works propose the large Galactic magnetic halo with very
intensive fields. The first article was published by \cite{ahn}, they
speculate about large and intensive purely azimuthal magnetic field in
the Galactic halo. This field should exist as an analogy to a solar wind
and should extend to about 1.5 Mpc. In spherical coordinates $r, \theta,
\phi$ it holds then
\begin{equation}
B_{\phi} = B_S R \frac{\sin \theta}{r},
\end{equation}
where $B_S R$ is the normalization factor derived from the values
in the solar surroundings, which is equal to 70 $\mu$G.kpc. If
such field is introduced, the positions of 11 out of 13 EHECRs
from Haverah Park, Volcano Ranch, Fly's Eye and AGASA should fall
within 20$^{\circ}$ spherical cap around M87 position. This
hypothesis was challenged shortly after its publication by
\cite{billoir}. They proved that this at the face-value exciting
fact, that M87 could be a single source of UHECRs, is simply based
on the fundamental property of the used magnetic field model in
halo. The used model of an azimuthal field is simply focusing all
positions into the direction of Galactic north pole and M87 is
lying near to this pole, and so the small angular distance between
computed EHECR positions and between M87 is probably just an
interesting coincidence without fundamental physical importance.
Furthermore, such strong magnetic halo is in contradiction with
recent observations.

\begin{figure}
\resizebox{\hsize}{!}{\includegraphics{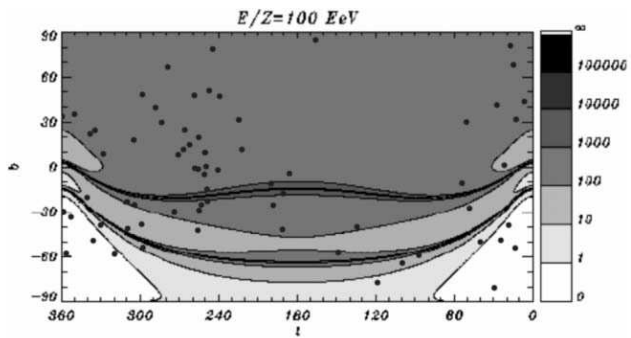}}
\caption[ ]{\em  Contour plots of the amplification of cosmic-ray flux
by the Galactic wind (according to \cite{harariold}). The plotted
dependance of the flux on arrival directions to the Earth was computed for
initial isotropic distribution of point sources (outside of the Galaxy) and
for $\bar E = 10^{20}$ eV. Figure was superposed with
coordinates of SUGAR events {\em (black dots)} of the arrivals of most energetic
particles ($\ge 4 \times 10^{19}$ eV according to Hillas E model of
energy estimation of SUGAR events). There are some SUGAR events inside
of the white triangle-shaped areas (bottom corners), where the zero
cosmic-ray flux is expected.}
\label{harrari}
\end{figure}

The second work was published by \cite{harariold} and it proposes the
Galactic magnetic wind extending to 1.5 Mpc. The model examines focusing
abilities of magnetic wind. Model of
the magnetic wind used in this work is purely azimuthal:
\begin{equation}
B = B_7 \frac{r_0}{r} \sin \theta \tanh \left( \frac{r}{r_s} \right),
\end{equation}
It describes $B$ as a function of the radial spherical coordinate
$r$ and the angle to the north galactic pole $\theta$. The term
$r_0$ in this equation is the distance from the Earth to the
Galactic center (equal to 8.5 kpc), factor $r/r_s$ was introduced
to smooth out the field at small radii ($r_s$ was taken as 5 kpc).
$B_7$ is the normalization factor (the strength of the field in [7
$\mu$G] units) and so in conservative models of GMF $B_7$ should
be $\sim 0.3 \div 0.4$. As it is shown in our combined Fig.
\ref{harrari}, such magnetic field has to sweep out some fraction
of the southern Galactic hemisphere. However, using the data from
SUGAR\footnote{We note that the SUGAR direction measurements are
generally significantly more trusted than their energy estimates.}
which are also plotted into this figure, we are able to show that
such a model could not be completely correct. This is due to the
fact that these regions with proposed zero particle flux --- in
contrary to the theoretical expectations
--- contain several SUGAR events.

Finally, two interesting works treating the turbulent fields
appeared recently. \cite{jaime} carefully analyzed the influence of turbulent
fields on possible clustering of UHECRs and \cite{harari} made large
study of properties of typical turbulent fields with respect to
amplification and multiplication of source images.

\subsection{Computer Model}\label{compres}

In our simulation we have supposed the conservative Galactic
magnetic field model by \cite{han}, which was amended with
toroidal and poloidal field components and with turbulent fields
linked to spiral arms; this complete configuration of magnetic
field was discussed in detail above.

\begin{figure}
\resizebox{\hsize}{!}{\includegraphics{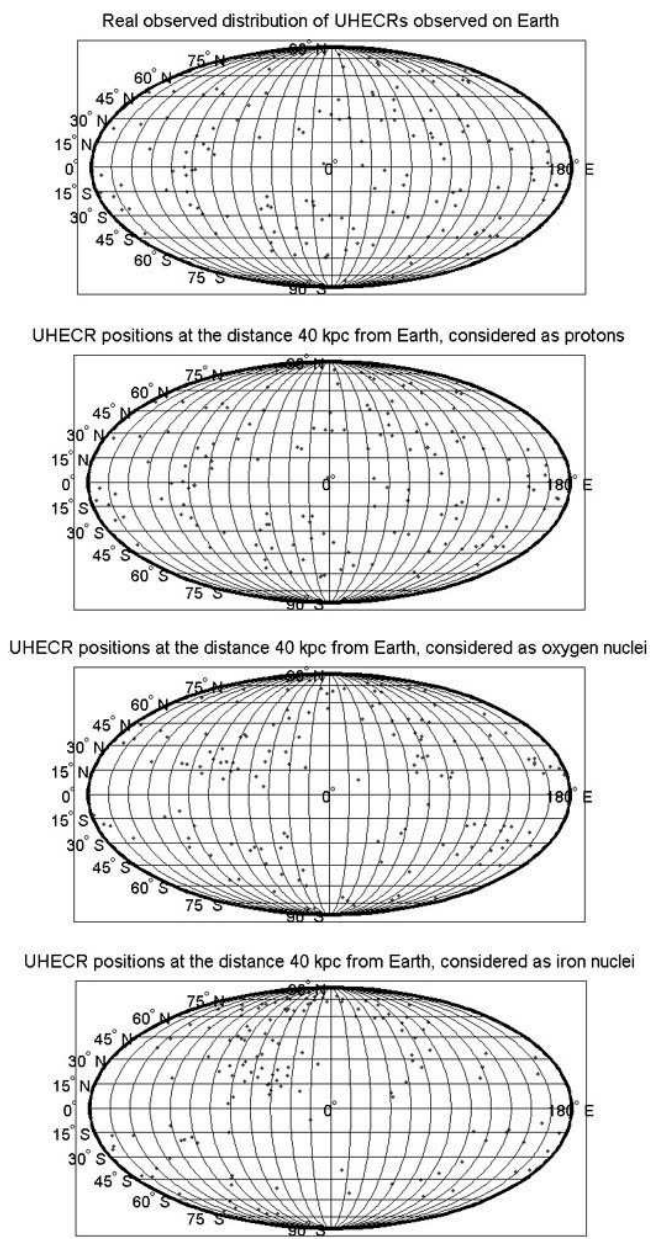}} \caption[
]{\em The original arrival directions (in galactic coordinates) of
145 UHECRs are shown in the uppermost part of the Figure. The
other three sky-maps (all in galactic coordinates) are showing the
final positions of cosmic ray particles which were propagated to
the distance 40 kpc from the Earth. The second map from the top
shows final positions of protons (the average angle between the
initial and final position of individual particle is $2.9^{\circ}
\pm 1.6^{\circ}$ [s.e.]). The third map shows final positions of
oxygen nuclei (average angle $23^{\circ} \pm 12^{\circ}$ [s.e.])
and finally the fourth map shows positions of iron nuclei (average
angle $72^{\circ} \pm 38^{\circ}$ [s.e.]).} \label{finalfig}
\end{figure}

Despite using various types of initial data, we present here only
the results for real data. Namely, even such constrained set of
data can sufficiently demonstrate all important changes of
features of particle flux. These real data\footnote{The arrival
direction ($b, l$) and energy $E$ was used for each detected
particle.} were taken from our catalogue of UHECRs\footnote{This
catalogue was created using available data from several various
experiments: Data for from all UHECR experiments with energies
above $10^{20}$ eV, data from AGASA experiment and data from SUGAR
experiment for particles with energies above $4 \times 10^{19}$ eV
were used. The catalogue is available on-line
(http://www-hep2.fzu.cz/Auger/catalogue.html).}. We propagate
these particles through Galactic magnetic field assuming various
charges --- starting as protons (proton number $Z = 1$),
continuing as oxygen nuclei ($Z = 8$) and ending with iron nuclei
($Z = 26$). All particles were traced back off the influence of
Galactic field. The final distance of each particle was assumed to
be 40 kpc from Earth. We present here (Figure \ref{finalfig}) the
results of simulations corresponding the real UHECR data (145
UHECR positions and energies) taken successively as protons,
oxygen and iron nuclei.

\subsection{Results of particle tracking}

We can state that the given deflection ranges (Fig. \ref{finalfig}) are
in good agreement with previous models (\cite{stanev1},
\cite{medina} or \cite{olinto}) of propagation of UHECRs through the
Galactic magnetic field. Hence we can formulate the following conclusions:
\begin{itemize}
\item{As we already stressed above, the detail of global structure
of GMF is still \textbf{uncertain}, but despite that we can claim
that its influence is non-negligible for protons and essential for
Fe nuclei.} \item{The simulations of particles with higher charges
(e.g. oxygen or iron nuclei) are transforming the isotropic
distribution to structures, which show some regularities. The
actual forms of these regular structures are as well as the global
model of GMF rather uncertain, but their existence could be taken
for granted and it is independent on the specific parameters of
given magnetic field model\footnote{Of course, our simulation does
not completely exclude the possibility that also the initial
directions of particles before they enter into the Galaxy are
isotropic. Our conclusions were derived only in one direction of
implication --- the observed isotropic distribution doesn't
necessarily require the initial isotropic distribution for oxygen
and iron nuclei. For test of opposite direction of implication we
have to make another type of simulations --- we have to inject
huge numbers of particles isotropically distributed on spherical
surface around Galaxy and then detect them on some tiny sphere (or
other shape) around Earth's position. This problem was partially
treated by \cite{olinto}.}. In accordance with \cite{harari} we
observe especially for oxygen and iron nuclei (Fig.
\ref{finalfig}) that at some places the initial flux is amplified,
in other areas it is strongly suppressed (see e.g. overdensity in
region to north-west from Galactic center for both oxygen and iron
nuclei or almost empty region along Galactic plane again for both
type of nuclei).}

\item{GMF is very important also for protons, because it is able to
affect the small-angle clustering (as one can see on the second upper
part of Fig. \ref{finalfig}, where some initial small clusters were transformed
into other ones). Small-angle clustering is today lively discussed and
it is one of the key features in discrimination between some models of
sources (\cite{jaime}).}

\item{The possibility that the UHECRs originate in the Galaxy
(e.g. near the young neutron stars in the form of iron nuclei) is
not probable, but not completely excluded (see the bottom part of
Fig. \ref{finalfig}). Furthermore, such UHECRs should originate
only in several point sources in our Galaxy, what is again in
accordance with the existence of pseudo-regular structures after
propagation through the GMF (see also \cite{olinto} or
\cite{harari}).}
\end{itemize}

The theory of Galactic origin of UHECRs could be also combined
with the above discussed fact that also relatively strong ($\sim
1$ mG) fields exist in the form of filaments near Galactic center.
In such field the Larmor radius of $10^{19}$ eV UHECR proton is
only about 4 pc.

\section{Application No.2: Chemical composition of CRs}
\subsection{Propagation of CRs in our model}

We used very simple method to model the propagation of cosmic ray
particles in a rather wide range of energy ($10^{13} \div 10^{19}$
eV). The model of regular magnetic field described above was
improved with following configuration of turbulent components:

The Galaxy was divided into cubic cells of an assumed size $L$.
Two values of cell length were studied, in particular 10 and 50
pc. The random orientation and strength of the turbulent magnetic
field were generated in given fraction of cells and also their
positions were random. In accordance with observation the
contribution of the turbulent magnetic field was taken equal to
$(0\div3)\times B(r,\theta,z)$, where $B(r,\theta,z)$ is sum of
strengths of the non-turbulent components. We neglected all
possible interactions of particles with matter and we kept the
energy of particles constant.

Our Galaxy model has the following geometrical boundary: the bulge
is a symmetric ellipsoid with a major axis in the Galactic
midplane 3 kpc long and a minor axis of 2 kpc. Around the bulge
there is a thin cylinder with a radius of 15 kpc and the height of
500 pc. Starting positions of particles reside in the Galactic
plane inside an annulus with radii of 3 and 12 kpc. This
assumption is in the agreement with the observed positions of
supernovae remnants (\cite{green}), which are the most probable
sources of CRs below the knee in our Galaxy\footnote{The density
of SNRs is higher in the bulge, but we have been interested in how
CRs behave in Galactic disk, where it is possible to compare our
results with the observations.}.

As \cite{gaisser00} and \cite{brunetti} have shown, the average
time spent in the Galaxy by cosmic ray of energy within the range
$(1\div100)\times10^9$ eV is $\tau\sim 10^{14}$ s. The energy
dependence of $\tau$ can be measured by comparing the spectrum of
the secondary nuclei to that of the parent primary nuclei. From
observation one can deduce that at least in the range
$10^{10}\div10^{12}$ eV, the mean residence time varies
approximately as $R^{-0.6}$ (\cite{garcia}, \cite{swordy} and
\cite{engelmann}), where $R=\frac{pc}{Ze}$ is the rigidity of
particle with momentum $p$ and atomic number $Z$. This
extrapolation breaks down around $3\times10^{15}$ eV
(\cite{gaisser00}) because the value of an effective escape length
is equal to $c\tau\sim300$ pc which corresponds to just one
crossing of the Galactic disk and the probability of nuclei escape
significantly rises. The situation within the highest energy range
is not clear, but we expect that the nuclei are not trapped in
GMF. The task we have to solve is to find the value of tracking
time for the simulation of particles with energies in range
$10^{13}\div10^{19}$ eV. We have found that the value
$T=3\times10^{12}$ s $\sim10^5$ yr appears as the most suitable
tracking time of particles for the study of nuclei escape rate
from the Galaxy. From the equation $\tau\sim R^{-0.6}$ we obtain
the value $10^{11}$ s for proton with the energy in the middle of
our range (which is equal to $10^{15}$ eV). Despite of it, we use
tracking time longer by one order of magnitude. The reason for
such choice is that: (1)~The mean residence time for nuclei with
higher $Z$ will be longer than for proton. (2)~The nuclei escape
rates are too high (too low) for longer (shorter) tracking times
and as such they are not suitable for discrimination between the
different nuclei. (We note that we use only one value of tracking
time for whole energy range of particles.)

The propagation of particle was stopped in a moment when the
particle escaped from the Galaxy. The escape occurred when the
particle crossed the Galaxy geometrical boundary. Otherwise, if
the particle stayed within the Galaxy for time longer than
$T=3\times10^{12}$, simulation was also stopped and the particle
was simply taken as not escaped. From these values of the particle
escape rates one can easily calculate the chemical composition of
CRs.

Our starting chemical composition is taken from \cite{wiebel}, who summarized
results of several experiments for energy $10^{12}$ eV. We have divided all nuclei
into five groups according to their mass. From each group we choose a nucleus that
is the best representative. In this way we have chosen protons and nuclei of helium,
oxygen, magnesium and iron as group representatives, with initial abundance equal to
$42\%$, $26\%$, $13\%$, $9\%$ and $10\%$, respectively. As
the indicator of the composition, we use mean value of the
logarithm of mass number $A$,
\begin{equation}
<\ln (A)>=\frac{\sum n_i (\ln (A_i))}{\sum n_i},
\end{equation}
where $n_i$ denotes the number of elements $i$ with mass number
$A_i$. The initial composition at $10^{12}$ eV  is $<\ln(A)> =
1.41$ (\cite{wiebel}).

\begin{figure}
\resizebox{\hsize}{!}{\includegraphics{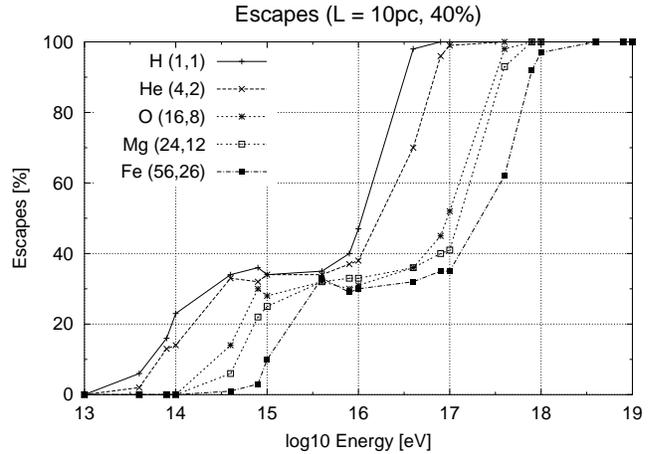}} \caption[ ]{
An example of nuclei escape rate from GMF improved by turbulent
MFs located in 40 percents of cubic cells with size equal to 10
pc.}\label{radek2}
\end{figure}

\subsection{Results and conclusions}

We have used only one above discussed model of GMF in our
simulation, although it was improved by random components of the
turbulent magnetic field. We have confirmed the influence of such
turbulent magnetic fields on the propagation of CRs for all
studied nuclei energies.

\begin{figure}
\resizebox{\hsize}{!}{\includegraphics{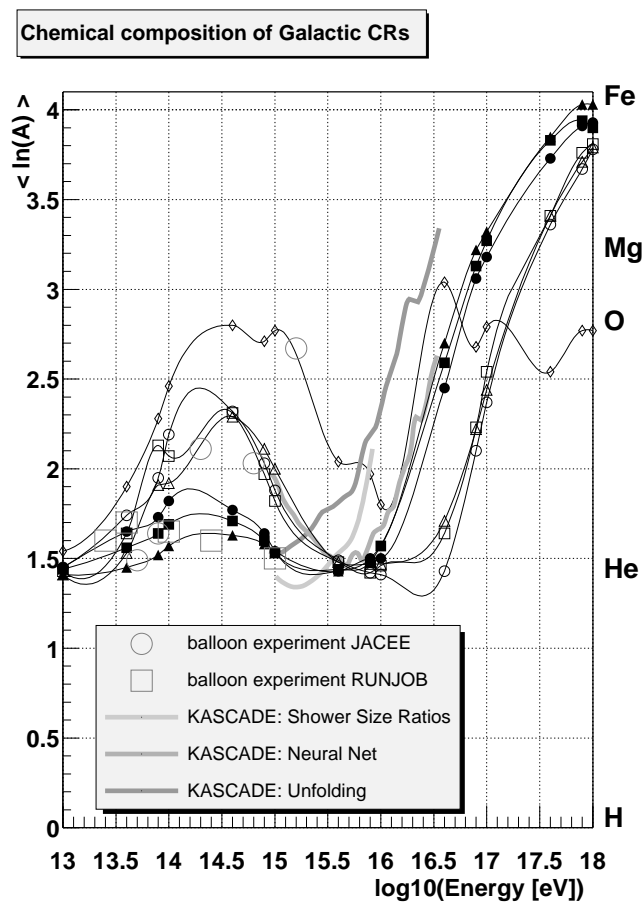}} \caption[ ]{
Change of nuclei composition as a result of nuclei escapes from
Galactic disk compared with experimental results. The line with
diamonds correspond to modelling without turbulent magnetic field,
the lines with circles, squares and triangles indicate number of
cells with turbulent magnetic field equal to 10 \%, 20 \% and 40
\% respectively; filled for dimension of cells equal to 50 pc and
empty for 10 pc.}\label{radek1}
\end{figure}

We find following results in our simulation:
\begin{itemize}

\item{The dependence of nuclei escape rate on the energy is
similar for all configurations of magnetic fields (Fig.
\ref{radek2}). Except configuration without turbulent magnetic
field all values of nuclei escape rate are lower than 7\% at our
starting energy $10^{13}$ eV (even for protons). Thereafter up to
$10^{15}$ eV the leakage depends on the charge, the higher charge,
the lower is number of nuclei escape rate. In the energy range
$10^{15}\div10^{16}$ eV the nuclei escape rate of light nuclei (H,
He) becomes constant and the values of heavier elements come
closer to them. The differences between cells with dimension equal
to 10 pc are 10\%, the lowest value is for the configuration with
the highest rate of the cells with turbulent magnetic field. The
situation for the cell dimensions equal to 50 pc is similar but
the values are much closer and lies around $(78\pm 2)$\%. Nuclei
with energies higher than $10^{16}$ eV behave in the same way as
at the energy below $10^{15}$ eV. It is again a function of
particle charge and we can observe an increase of the abundance of
heavy elements (Fig. \ref{radek1}). Escape rate achieves 100\% for
protons (all protons leave the Galaxy) at energy equal to
$4\times10^{17}$ eV. Protons are closely followed by nuclei of
helium and for energies higher than $10^{18}$ eV no particle will
remain in Galactic disk.}

\item{We have found that more than 90\% of the particles above
$10^{16}$ eV are escaping independently of their charge from the
regular magnetic field. We believe that the different nuclei
escape rate for the energies above $10^{16}$ eV is caused mainly
by random component of GMF.}

\item{The leakage of nuclei from the Galaxy depends significantly
on the characteristics of turbulent magnetic fields (field's
strength, their dimensions and locations in the Galaxy and also on
the number of cells with turbulent magnetic field). It follows
from our simulations that the higher the fraction of the cells
with turbulent magnetic fields, the slower is the leakage. This is
because of the nuclei are trapped in these cells and their leakage
from the Galaxy decreases. Unfortunately all properties of
turbulent magnetic fields, which are very important for the
propagation of CRs, are not known enough.}

\item{The behavior of protons and helium nuclei is very similar in
the whole studied energy range. They escape more easily than
nuclei with higher charge (oxygen, magnesium and iron). Despite of
this, they still play important role in CRs, because of their
dominant abundance in the initial composition of CRs (representing
together more than two thirds of all particles).}

\item{The result of the different nuclei escape rate is the
increase of the abundance of heavier nuclei in the chemical
composition of CRs. The comparison of the chemical composition
resulting from our modelling with the measurements\footnote{We
compare our results only with few experiments, the full review of
other can be found in \cite{wiebel} and references therein.} is
shown in Fig. \ref{radek1}. Experimental data show unique results
only for the energies below $10^{14}$ eV, where they show slight
increase of mean $\ln(A)$. Above $10^{14}$ eV the results of two
balloon experiments JACEE (\cite{jacee}) and RUNJOB
(\cite{runjob}) disagree. RUNJOB shows no change in the chemical
composition (constant value of mean $\ln(A)$), whereas data from
JACEE indicate increase of mean $\ln(A)$. The experimental results
from RUNJOB are in agreement with our results --- for our model
with turbulent magnetic fields with cell dimensions equal to 10
pc. In the case with larger dimension of cells (50 pc) we have
found higher leakage of light nuclei (proton and helium) which
leads to the same increase of mean $\ln(A)$ as measured by JACEE.

For energies above $10^{15}$ eV we have chosen only data from
KASCADE (\cite{haungs} and \cite{horandel})\footnote{We must note
that the chemical composition above $10^{15}$ eV is not clear, the
results from different experiments do not agree and there is also
problem with reconstruction of extensive air showers, which leads
to different determination of chemical composition detected in one
experiment.}. Two methods of data processing show increasing mean
$\ln(A)$ from the value equal to the initial composition at
$10^{12}$ eV to the value, when the majority of cosmic rays is
composed of heavy elements. Third method (neural net) gives
different results, firstly decrease from the value $ <\ln(A)> = 2$
to the initial value and above energy equal to $10^{16}$ eV
increase to high value of mean ln(A).

The results of our modelling have following characteristics above
$10^{15}$ eV: The modelling with only regular GMF does not agree
with experimental data, while the cases with turbulent MF show
correctly the increase of the abundance of heavy elements at high
energies. The discrepancy between our model with turbulent MF and
measurements is in the value of the energy, where the increase of
mean $\ln(A)$ starts. If we take the energy equal to
$3\times10^{15}$ eV (knee) as correct, then the differences will
be half magnitude and one magnitude for dimensions of cells equal
to 10 and 50 pc, respectively. We have found that this discrepancy
does not strongly depend on general parameters of our modelling
(the tracking time and thickness of Galactic disk), we believe
that this discrepancy can be removed only using more realistic
method of modelling motion of atomic nuclei (for review see
\cite{gaisser90}).}

\item{The different leakage of nuclei from GMF produces a break at
energy $10^{16}$ eV in our modelling, similar to observed
characteristic of cosmic ray flux, known as the knee
($3\times10^{15}$ eV). Thus our simple model of the propagation of
Galactic cosmic rays favors theory about origin of the knee
presented by \cite{ptuskin}.}

\item{The model of propagation in GMF without turbulent MF seems
to us as very unrealistic, because there is in strong disagreement
with measurements.}

\item{We can see that the chemical composition depends on the
characteristics of turbulent MF, so it gives us possibility to
deduce these characteristics from the abundance of elements in
Galactic cosmic rays.}
\end{itemize}

The used method is a simple way to simulate the propagation of CR
within wide energy range. Despite of good obtained results we
conclude that the propagation of particles must be solved by more
realistic method, especially for particles with energies below
$10^{16}$ eV. We used only one type of the source in the Galactic
midplane with constant chemical composition. However, there are
indications that we must expect more types of sources in the
Galactic and extragalactic space resulting in more complicated
cosmic ray flux.

\begin{acknowledgements}
We would like to thank to Ji\v{r}\'{\i} Grygar and to Jan
\v{R}\'{\i}dk\'y for the grand help with the origin of this paper.
We also thank to Petr Harmanec and to Jan Palou\v{s} for their
valuable comments, which helped us to generally improve the
structure of this paper. We also would like to thank to Jin Lin
Han for his great advices and consultations about structure of
Galactic magnetic field and to referee for important suggestions,
which changed this paper to its present form. This work was
supported by the Grant No. A1010928/1999 of Grant Agency of the
Academy of Sciences of the Czech Republic, by the Grant LN00A006
of Ministry of Education through Center of Particle Physics and by
the Grant LA134 of Ministry of Education through Project INGO.
\end{acknowledgements}


\begin{thebibliography}{}
\bibitem[Ahn et al. (1999)]{ahn}
Ahn E.-J., Medina-Tanco G., Biermann P., Stanev T., 1999, Preprint
astro-ph/9911123 (submitted to Physical Review Letters)

\bibitem[Alvarez-Mu\~{n}iz et al. (2002)]{jaime}
Alvarez-Mu\~{n}iz J., Engel R. \& Stanev T., 2002, \apj 572, 185

\bibitem[Apanasenko et al. (2001)]{runjob}
Apanasenko A.V. et al., 2001, Astroparticle Physics 16, 13

\bibitem[Axford (1994)]{axford}
Axford W.I., 1994, \apss 90, 937

\bibitem[Beck (2001)]{beck}
Beck R., 2001, Space Science Reviews 99 (1), 243

\bibitem[Beck et al. (1996)]{beckold}
Beck R., Brandenburg A., Moss D., Shukurov A. \& Sokoloff D., 1996,
\araa 34, 155

\bibitem[Billoir \& Letessier-Selvon (2000)]{billoir}
Billoir P., Letessier-Selvon A., 2000, Preprint astro-ph/0001427

\bibitem[Brunetti \& Codino (2000)]{brunetti}
Brunetti M.T., Codino A., 2000, \apj 528, 789

\bibitem[Elstner et al. (1992)]{elstner}
Elstner D., Meinel R., Beck R., 1992, \aas 94, 587

\bibitem[Engelmann et al. (1990)]{engelmann}
Engelmann J.J., Ferrando P., Soutoul A., Goret P., Juliusson E.,
1990, \aaa 223, 96

\bibitem[Gaisser (1990)]{gaisser90}
Gaisser T.K., 1990, Cosmic Rays and Particle Physics, Cambridge
University Press

\bibitem[Gaisser (2001)]{gaisser00}
Gaisser T.K., 2001, AIP Conference Proceedings 558, 27

\bibitem[Garcia-Munoz et al. (1987)]{garcia}
Garcia-Munoz M. et al., 1987, \apjs 64, 269

\bibitem[Green (2001)]{green}
Green D.A., 2001, A Catalogue of Galactic Supernova Remnants,
http://www.mrao.cam.ac.uk/surveys/snrs/

\bibitem[Greisen, 1966]{greisen}
Greisen K., 1966, Physical Review Letters 16, 748

\bibitem[Han \& Qiao (1994)]{hanold}
Han J.L., Qiao G.J., 1994, \aaa 288, 759

\bibitem[Han et al. (1999)]{han}
Han J.L., Manchester R.N., Qiao G.J., 1999, \mn 306, 371

\bibitem[Han (2002)]{han_new}
Han J.L., 2002, AIP Conference Proceedings 609, 96

\bibitem[Han et al. (2002)]{han_norma}
Han J.L., Manchester R.N., Lyne A.G. \& Qiao G.J., 2002, \apj 570,
L17

\bibitem[Harari et al. (2000)]{harariold}
Harari D., Mollerach S., Roulet E., 2000, AIP Conference Proceedings
566, 289

\bibitem[Harari et al. (2002)]{harari}
Harari D., Mollerach S., Roulet E. \& S\'anchez F., 2002, The Journal of
High Energy Physics 03, 045

\bibitem[Haungs (2002)]{haungs}
Haungs A, 2002, Preprint astro-ph/0212481 v1, submitted to
J. Phys. G: Nucl. Part. Phys.

\bibitem[H\"{o}randel (2002)]{horandel}
Horandel J.R., 2002, Preprint astro-ph/0210453 v1, accepted for
publication in Astroparticle Physics

\bibitem[Hiltner (1949)]{hiltner}
Hiltner W.A., 1949, \apj 109, 471

\bibitem[Lee \& Clay (1995)]{lee}
Lee A.A., Clay R.W., 1995, J. Phys. G 21, 1743

\bibitem[Lemoine et al. (1997)]{lemoine}
Lemoine M., Sigl G., Olinto A.V., Schramm D., 1997, \apj 486, L115

\bibitem[Medina Tanco et al. (1998)]{medina}
Medina Tanco G.A., de Gouveia dal Pino E.M., Horvath J.E., 1998, \apj
492, 200

\bibitem[O'Neill et al. (2001)]
{olinto} O'Neill S., Olinto A., Blasi P., Proceedings of ICRC 2001,
Section OG 1.3, Paper No. 6890 (also Preprint astro-ph/0108401)

\bibitem[Ptuskin et al. (1993)]{ptuskin} Ptuskin V.S. et al., 1993,
\aaa 268, 726

\bibitem[Rand \& Kulkarni (1989)]{rand}
Rand R.J., Kulkarni S.R., 1989, \apj 343, 760

\bibitem[Swordy et al.(1990)]{swordy}
Swordy S.P. et al., 1990, \apj 349, 625

\bibitem[Stanev (1997)]{stanev1}
Stanev T., 1997, \apj 479, 290

\bibitem[Takahashi (1998)]{jacee}
Takahashi, 1998, Nuclear Physics B (Proceedings Supplement) 60B,
83

\bibitem[Tinyakov \& Tkachev (2001)]{tinyakov}
Tinyakov P.G. \& Tkachev I.I., 2002, Astroparticle Physics 18, 165

\bibitem[Wainscoat et al. (1992)]{wainscoat}
Wainscoat R.J., Cohen M., Volk K., Walker H.J., \& Schwartz D.E.,
1992, \apjs 83, 111

\bibitem[Wiebel-Sooth et al. (1998)]{wiebel}
Wiebel-Sooth B., Biermann P.L. \& Meyer H., 1994, \aaa 330, 389

\bibitem[Widrow (2002)]{widrow}
Widrow L.M., Preprint astro-ph/0207240 v1, accepted for publication in
Reviews of Modern Physics

\bibitem[Zatsepin \& Kuzmin, 1966]{zacepin}
Zatsepin G.T. \& Kuzmin V.A., 1966, Zh. Eksp. Theor. Fiz. (Pisma Red.)
4, 114
\end{thebibliography}
\end{document}